# Leveraging Convolutional Neural Networks for 3D Quantitative Angiography Reconstructions from Sparse Cone Beam CT Projections Utilizing CFD Data

Ahmad Rahmatpour[1,2], Allison Shields[1,2], Parmita Mondal[1,2], Parisa Naghdi[1,2], Michael Udin[1,2], Kyle A Williams[1,2], Mohammad Mahdi Shiraz Bhurwani[3], Swetadri Vasan Setlur Nagesh[1,2] and Ciprian N Ionita[1,2,3]

[1]Department of Biomedical Engineering, University at Buffalo, Buffalo, NY 14260
[2]Canon Stroke and Vascular Research Center, Buffalo, NY 14203
[3]QAS.AI Inc, Buffalo, NY 14203

## ABSTRACT

**Purpose:** This study leverages convolutional neural networks (CNNs) to enhance the temporal resolution of 3D angiography in intracranial aneurysms (IAs), focusing on the reconstruction of volumetric contrast data from sparse and limited projections.

**Materials and Methods:** Three patient-specific IA geometries were segmented and converted into stereolithography files to facilitate computational fluid dynamics (CFD) simulations. These simulations first modeled blood flow under steady conditions with varying inlet velocities: 0.25 m/s, 0.35 m/s, and 0.45 m/s. Subsequently, 3D angiograms were simulated by labeling inlet particles to represent contrast bolus injections over durations of 0.5s, 1.0s, 1.5s, and 2.0s. The angiographic simulations were then used within a simulated cone beam C-arm CT system to generate in-silico rotational DSAs, capturing projections every 10 ms over a 220-degree arc at 27 frames per second. From these simulations, both fully sampled (108 projections) and truncated projection datasets were generated—the latter using a maximum of 49 projections. High-fidelity volumetric images were reconstructed using a Parker-weighted Feldkamp-Davis-Kress algorithm. A modified U-Net CNN was subsequently trained on these datasets to reconstruct 3D angiographic volumes from the truncated projections. The network incorporated multiple convolutional layers with ReLU activations and Max pooling, complemented by upsampling and concatenation to preserve spatial detail. Model performance was evaluated using mean squared error (MSE).

**Results:** Evaluating our U-net model across the test set yielded a MSE of $10^{-4}$, indicating good agreement with ground truth reconstructions and demonstrating acceptable capabilities in capturing relevant transient angiographic features.

**Conclusions:** This study confirms the feasibility of using CNNs for reconstructing 3D angiographic images from truncated projections, effectively capturing transient angiographic features. Further validation and refinement are necessary to advance clinical implementation.

**Summary:** This study explores the use of convolutional neural networks (CNNs) to improve 3D quantitative angiography by reconstructing full 3D angiographic images from truncated projection data. Using patient-specific vascular geometries, 3D virtual angiograms were generated via computed fluid dynamics and captured with C-arm cone beam computed tomography. A modified U-Net CNN was trained on sparse reconstructions from 29 and 49 projections, achieving a mean square error of $10^{-4}$. Data augmentation enhanced the model's robustness. The results demonstrate the model's ability to accurately reconstruct angiographic data, offering promising potential for clinical applications in diagnosis and treatment planning.

## 1. BACKGROUND

Intracranial aneurysms (IAs) pose a significant risk of hemorrhagic stroke due to their potential to rupture, making timely and accurate diagnosis and treatment crucial. [1] The current gold standard for imaging aneurysms during endovascular treatment is digital subtraction angiography (DSA), which provides excellent spatial and temporal resolution but remains a 2D technology. [2] This limitation can lead to oversimplifications in the visualization and analysis of the complex 3D structures of cerebral vasculature, particularly in the Circle of Willis, where these aneurysms commonly occur. [3, 4]

While modern endovascular techniques have advanced the management of intracranial aneurysms, predicting procedural outcomes and planning treatments effectively require more detailed visualization than traditional 2D DSA can offer. Recent developments in machine learning technology provide a prognosis of the IA occlusion rates and assist in procedural planning. [5, 6] They show promise, but they could be further improved if 3D information could be provided in a reliably fashion. A more complete approach to quantifying the 3D hemodynamics within these vascular lesions could significantly enhance clinical outcomes.

The rapid contrast propagation through the arterial phase in the Circle of Willis, typically lasting 3-4 seconds, necessitates high-speed imaging capable of capturing dynamic changes within this brief window to effectively "freeze" the moment for detailed analysis. Achieving this level of detail in 3D would allow clinicians to extract time-dilution curves at every point in the artery, enhancing the understanding of aneurysm behavior and blood flow dynamics. Our study seeks to address this need by employing a convolutional neural network (CNN) to reconstruct dynamic 3D angiographic volumes from limited projection data, utilizing fewer than the traditionally required number of projections. [7] This research builds on previous successes in reconstructing vascular geometries with minimal projections through segmentation techniques. We now aim to extend this to attempt capturing CT-like volumetric information with improved temporal resolution, to extract the 3D and temporal contrast gradients during the bolus transit through the complex arterial networks around IAs. By generating high-fidelity computational fluid dynamics (CFD) simulations and mimicking C-arm CT acquisitions, we investigate the degree of projection truncation acceptable for producing clinically viable 3D quantitative angiography (3D-QA), pushing the boundaries of current imaging capabilities for better diagnostic and therapeutic interventions in intracranial aneurysm management.

## 2. MATERIALS AND METHODS

### 2.1 Data Generation

This study was conducted with approval from the State University of New York at Buffalo institutional review board. We acquired 3D patient-specific CT angiography data from three patients with intracranial aneurysms. Each vascular geometry was transformed into a stereolithography (STL) file format, followed by simplification of the distal arterial geometry to facilitate CFD simulations. The flow domain was meshed using ANSYS ICEM to ensure precise resolution and boundary adherence. We simulated the blood flow under steady flow conditions, specifying inlet velocities of 0.25 m/s, 0.35 m/s, and 0.45 m/s. For these simulations, we introduced labeled particles at the inlet to mimic the transit of a contrast bolus, with injection durations set at 0.5s, 1.0s, 1.5s, and 2.0s, respectively. To capture the dynamic progression of the contrast bolus through the vascular network, data was recorded every 10 ms, generating thousands of volumetric angiograms for each case. These angiograms served as a detailed temporal series, capturing the bolus movement, essential for subsequent imaging reconstructions.

For image acquisition simulations, we simulated a cone beam C-arm CT (CBCT) system, Figure 1. This system performed a full rotation at 27 frames per second, capturing 108 projections over a 220-degree arc with a 25% duty cycle. The synthesized imaging data included both fully sampled and truncated acquisition (i.e. less than 220 degrees). The full CBCT datasets were reconstructed using a Parker-weighted Feldkamp-Davis-Kress (FDK) algorithm, providing high-fidelity volumetric images. Additionally, truncated datasets were generated to simulate conditions of incomplete data acquisition, as typically encountered in clinical settings.

This methodology enabled the creation of detailed 3D-QAs, with both fully reconstructed and truncated reconstruction datasets which were subsequently used to train a modified U-Net CNN.

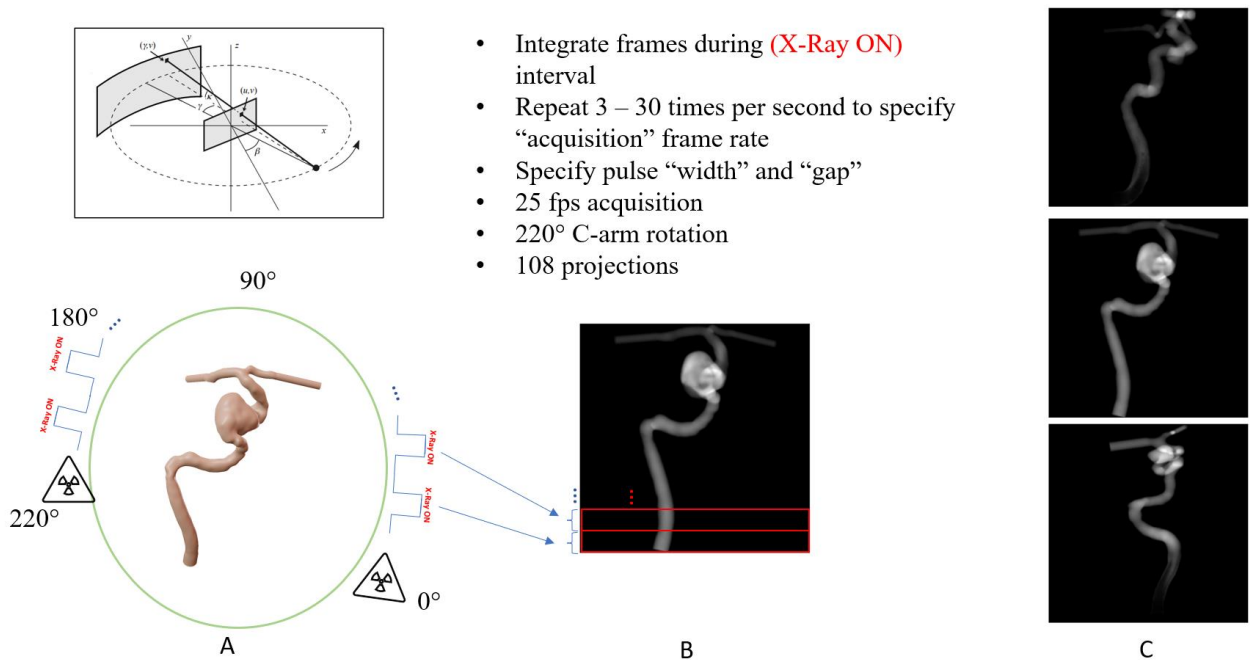

Figure 1. A) Simulate Cone beam CT. B) Generation of phantoms. C) 3 different phantoms used to train and validate the convolutional neural network.

## 2.2 Model investigation

A CNN was developed and tested to reduce truncation artifacts in CBCT reconstruction, which can impede clinical diagnostics. The model operates on input images of size 256×256 pixels with 256 slices and with a single channel. The network architecture utilizes a series of convolutional, activation, and pooling layers to encode the input images into a reduced dimensional representation, followed by a symmetrical decoding pathway to reconstruct the artifact-reduced images, Figure 2. Specifically, the encoding path consists of three convolutional blocks, each comprising two convolutional layers with ReLU activation and a subsequent max pooling operation for spatial downscaling. The convolutional layers in these blocks use kernel sizes of 8×8, 5×5, and 3×3, respectively, with the number of filters doubling in each subsequent block, starting from 64 filters in the first block.

At the bottleneck of the network, the feature maps are processed by two additional 3×3 convolutional layers with ReLU activation without any pooling, preparing the features for the decoding path. The decoding pathway mirrors the encoding steps but employs upsampling and concatenation operations with corresponding feature maps from the encoding path to preserve spatial hierarchies and details. The final output of the network is obtained by passing the decoded features through a 1×1 convolutional layer with a linear activation function, ensuring the output image dimensions are identical to the input. For training, the model utilizes the Adam optimizer with a mean squared error loss function. The truncated reconstructions served as the training set, while the corresponding fully reconstructed CT slices were used as ground truth. TensorFlow's data handling utilities were employed to efficiently manage and process these datasets in batches. The training process is further supported by several key callbacks, including model checkpoints to save the best model weights, ReduceLROnPlateau to adjust the learning rate based on performance, and EarlyStopping to halt training if the model performance does not improve. The learning rate for the optimizer is set to 0.0001. The model's performance was evaluated using mean squared error (MSE), assessing the difference between the predicted and ground truth images.

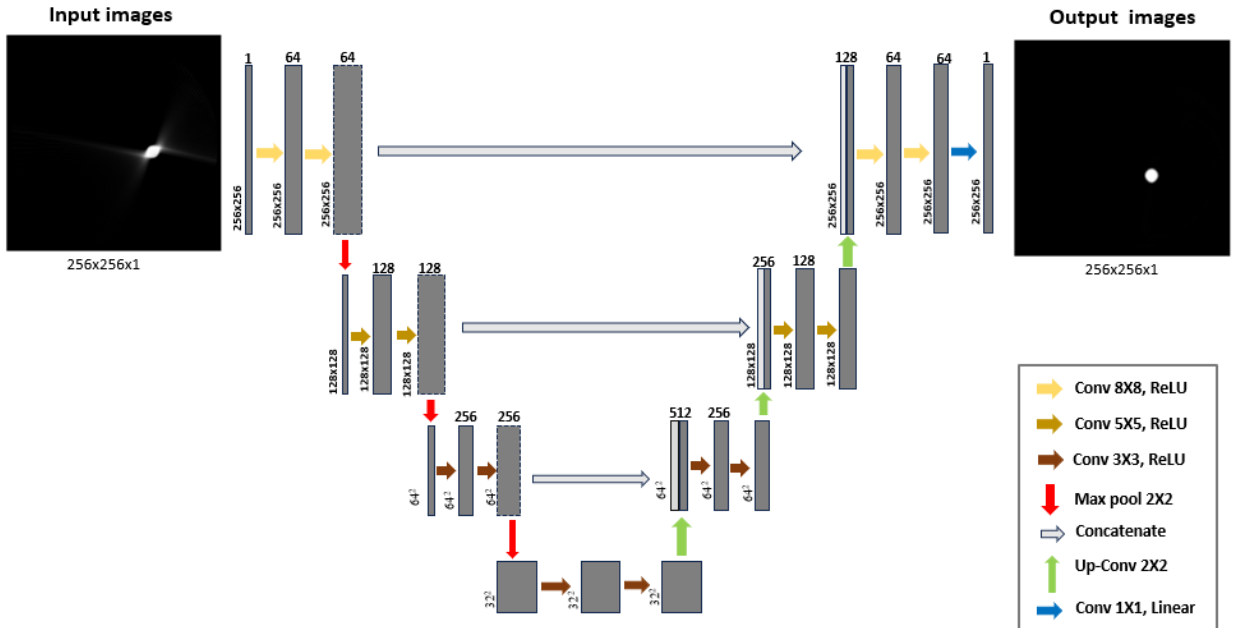

Figure 2. U-Net architecture. The network consists of a contracting path (left) for feature extraction and a symmetric expanding path (right) for precise localization. The contracting path downsamples the input image using convolutions and max pooling, while the expanding path upsamples the feature maps and combines them with corresponding features from the contracting path.

## 3. RESULTS

U-Net model effectively reconstructed full angle images from sparse angle projections of phantoms. The highest level of performance was attained when training and assessment concentrate on the angles that provide the most opacification, as evidenced by the comparison of ground truth phantoms and model predictions in Figure 3.

Validation of the U-Net model on the test set yielded a MSE of $10^{-4}$. This indicated that the model performed well in accurately reconstructing the ground truth and was capable of capturing important transient angiographic features, little variations in the contrast agent as it progressed through the vessel. The progression of this contrast agent over time was reconstructed using our U-Net model, Figure 4.

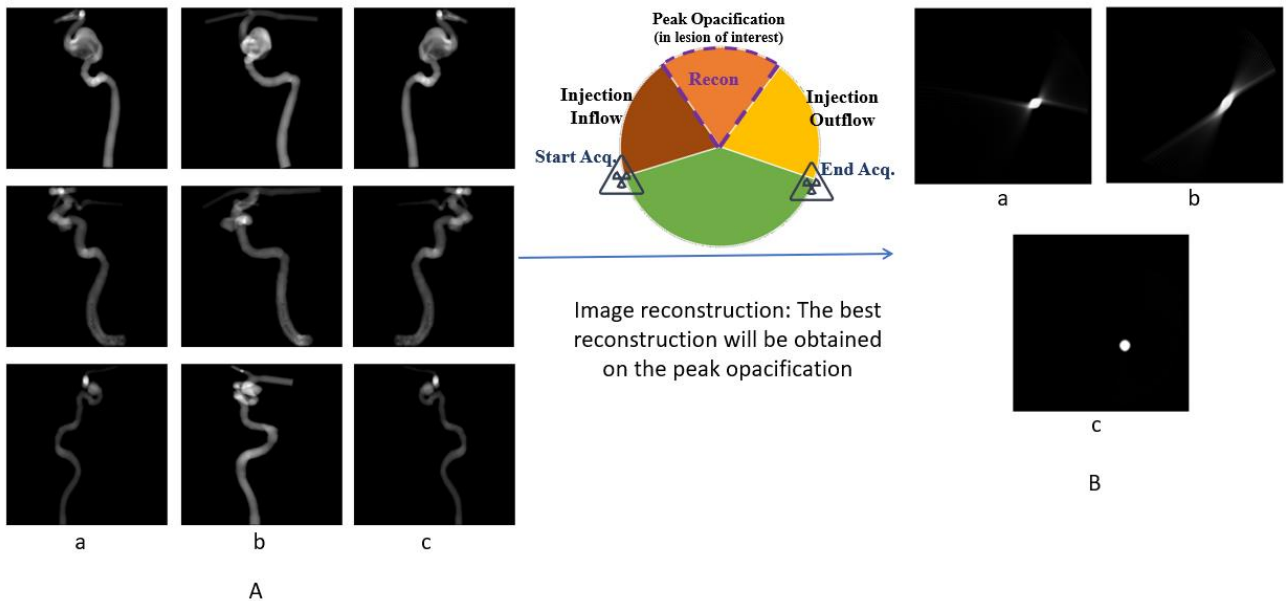

Figure 3. A) Phantoms generated by computational fluid dynamics from different angle projections, a) 0º, b) 90º and c) 180º. B) Truncated reconstructions: a) 100° and b) 60°, c) ground truth.

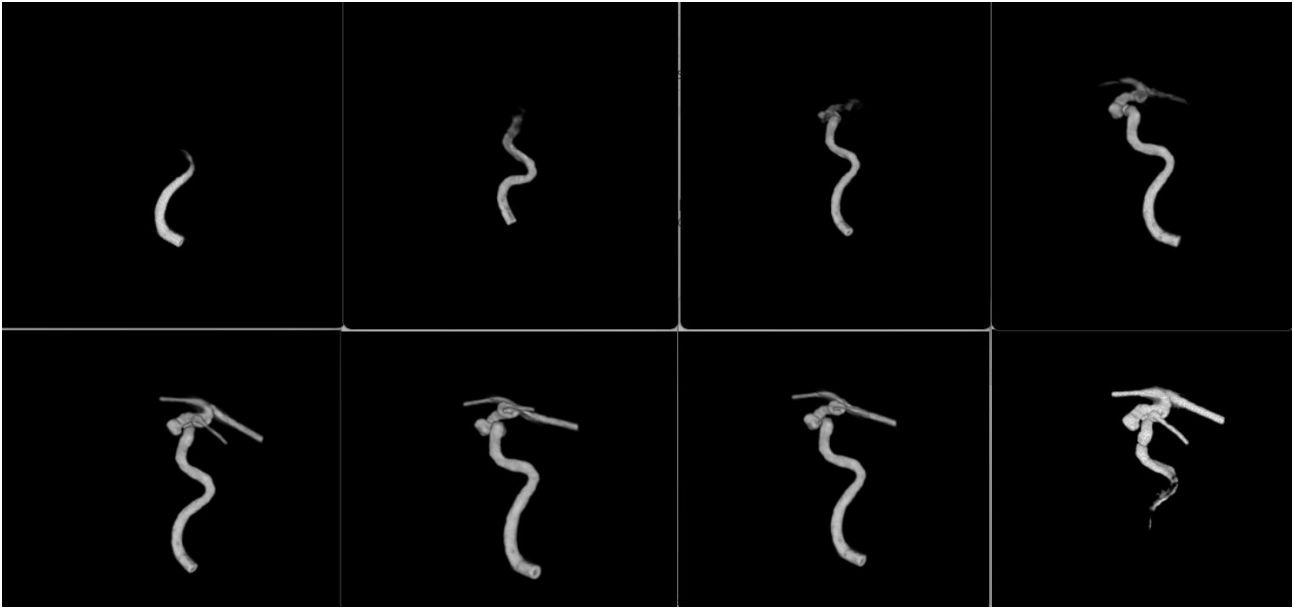

Figure 4. Reconstructed phantom images with varying contrast timing. The contrast agent was introduced into the vessel, and its progression is visualized over time.

## 4. CONCLUSION

This study demonstrates that leveraging a modified U-Net CNN for the reconstruction of 3D angiographic volumes from truncated projections enhances temporal resolution in intracranial aneurysm imaging. These advancements could lead to improved diagnostic capabilities and better clinical decision-making in the management of intracranial aneurysms.

## 5. ACKNOWLEDGMENTS

The findings from this study demonstrate the feasibility of using CNNs to reconstruct dynamic 3D angiographic data using truncated projection sets. While the results are promising, showing accurate reconstruction of complex vascular geometries, this method primarily establishes a foundational approach for enhancing dynamic 3D visualization of IAs. The successful application in a controlled setting lays the groundwork for future studies aimed at translating these advanced imaging techniques into clinical practice, potentially improving diagnostic accuracy and treatment planning.